# A Fresh Look at FAIR for Research Software

Daniel S. Katz, Morane Gruenpeter, Tom Honeyman, Lorraine Hwang, Mark D. Wilkinson, Vanessa Sochat, Hartwig Anzt, Carole Goble, and FAIR4RS subgroup 1

9 February 2021

(v1.1, changes from v1.0: formatting in Appendix B)
(v1.2, changes from v1.1: principle R1 in §4 and Appendix B was incorrectly copied from §3)
(v1.3, changes from v1.2: minor spelling and grammar fixes, additions in Appendix B to better explain new principles)

*Abstract: This document captures the discussion and deliberation of the FAIR for Research Software (FAIR4RS) subgroup that took a fresh look at the applicability of the FAIR Guiding Principles for scientific data management and stewardship for research software. We discuss the vision of research software as ideally reproducible, open, usable, recognized, sustained and robust, and then review both the characteristic and practiced differences of research software and data. This vision and understanding of initial conditions serves as a backdrop for an attempt at translating and interpreting the guiding principles to more fully align with research software. We have found that many of the principles remained relatively intact as written, as long as considerable interpretation was provided. This was particularly the case for the "Findable" and "Accessible" foundational principles. We found that "Interoperability" and "Reusability" are particularly prone to a broad and sometimes opposing set of interpretations as written. We propose two new principles modeled on existing ones, and provide modified guiding text for these principles to help clarify our final interpretation. A series of gaps in translation were captured during this process, and these remain to be addressed. We finish with a consideration of where these translated principles fall short of the vision laid out in the opening.*

# 1. Introduction

We envision a world where
- All research is reproducible
- All research software is open
- All research software is usable by others (for their own research)
- All contributors to research software are recognized for their work
- All research software is sustained as long as it is useful
- All research software is high-quality and robust

Here, we are working to define a part of this vision, specifically, making research software Findable, Accessible, Interoperable, and Reusable (FAIR).

In May 2020, a group of researchers came together and proposed an open working group to create a community consensus about how the FAIR principles could be applied to software ([FAIR for Research Software, FAIR4RS](#)), and if changes were needed. While sponsors of this group (the [Research Data Alliance](#), the [Research Software Alliance](#), and [FORCE11](#)) were deciding about this proposal (which they approved in September 2020), the group decided to start working by creating subgroups to address various parts of the problems, and then held a pair of webinars and publicized its goals to obtain members. This paper is the output of subgroup 1, "A fresh look at FAIR for Research Software," which was charged to "examine the FAIR principles in the context of research software." Specifically, it aims to
1. Determine what part of the FAIR principles apply as is to research software;
2. Determine what part of the FAIR principles doesn't apply at all to research software; and
3. Determine what part of the FAIR principles applies to research software, but with a different definition or different details,

starting with the FAIR principles themselves, and not relying on work done by others to apply them to research software, such as by Lamprecht et al. (2020).

This document is not the deliverable of the FAIR for Research Software working group, though it is intended to be the basis for part of that deliverable. The full working group's deliverable will also include the results of three other subgroups, which will define research software, examine how FAIR is being applied to other types of non-data digital products, and examine a previous FAIR for research software paper (Lamprecht et al. 2020) and work that has used it.

We base our discussion on "The FAIR Guiding Principles for scientific data management and stewardship" (Wilkinson et al. 2016), commonly known and referred throughout as the FAIR principles. We worked specifically with the principles as listed by the GO FAIR initiative (https://www.go-fair.org/fair-principles/), also see Appendix A. These principles are divided into the "foundational principles" (Findable, Accessible, Interoperable, Reusable), and the additional "guiding principles" which further elaborate upon each foundational principle. We will examine both sets of principles.

GO FAIR says "The principles refer to three types of entities: data (or any digital object), metadata (information about that digital object), and infrastructure," and software can be considered a digital object. However, it is clear that this statement is not completely true: the principles, as written, do not directly translate to research software, particularly at a detailed level.

The goals of this document are to present the methodology used when defining the new potential FAIR principles for research software. Subgroup 1 worked by initially polling its members on their opinions of the 4 foundational principles and 15 guiding principles as applicable to research software as written, applicable but with changes, or not applicable. The group members then used a shared document to discuss each principle. Next, a draft of this summary document was written and shared with the subgroup for discussion and comments. Then, a subset of the subgroup who were willing to meet virtually had a series of about 6 calls over 2 months to further discuss and refine this document, leading to a semi-final draft, which was again shared with the full subgroup for comments over a two-week period. This final version includes changes made in response to those comments.

This document includes, in Section 2, a discussion of the differences between software and data, an initial straightforward translation that was collected from the FAIR4RS-subgroup1 participants, and in Section 3, a discussion about the nuances of the currently defined rules in context of research software. Finally, we propose a potential new set of principles for FAIR research software (in Section 4, and compared with the FAIR data principles in Appendix B), identify gaps in our current systems (in Section 5), and consider where the translated principles fall short of our overall vision (in Section 6).

## 2. Differences between software and data

There are many inherent differences between software and data, and additionally, there are differences between how they are created, maintained, and used in the scholarly ecosystem.

As discussed by Katz et al. (2016), software is data, but it is not just data. While "data" in computing and information science can refer to anything that can be processed by a computer, software is a special kind of data that can be a creative, executable tool that operates on data. Specific differences include:

- Software is executable, data is not.
- Data provides evidence, software provides a tool.
- Software is a creative work, scientific data are facts or observations.
- Software suffers from a different type of bit rot than data: It is frequently built to use other software, leading to complex dependencies on this software, and these dependent software packages also frequently change.
- Software (and scientific software especially) is sometimes highly optimised for the hardware and software environment on which it runs, making it far more dependent on changes to that environment, while data is more commonly expressed in a form abstracted away from these concerns.
- Data tends to have value for researchers over many decades or longer, while a particular piece of software is likely to be deprecated much sooner.
- Software, over its lifetime, is typically subject to many changes whereas data often is not.
- "Software" commonly refers to a complex composite of different objects such as source code and/or compiled binaries, data, and documentation, while data tends to be composed of fewer object types (if not only data).

The FAIR principles sought to bridge research and informatics practices, but the FAIR principles for research software brings in a third element that also must be considered: the well developed set of practices within software development. An additional hurdle experienced during discussion for this group is that many terms have very different meanings within these domains. Fundamental terms like data, metadata, repository, archive, publish, release, and identifier have very different meanings in the context of research data and in the context of research software. Combined, these present different baseline assumptions that must be considered when proposing a set of principles to bridge practice.

In addition to the inherent differences between software and data, there are differences in the way they are created, maintained and used in the scholarly ecosystem. The following contrasts in these areas are common (although there are exceptions on both sides). Differences between the two are **highlighted in bold**:

- Data may be more valued for its provenance and stability, while software may be more valued for its functionality, which may be iteratively developed and elaborated upon
- While there are areas of overlap, models of sustainability differ between the two, particularly because of differences in how they are treated as **intellectual property**
- Software is more likely to be used and cited while it is actively and iteratively developed, while the general expectation with data is that it is a **"finished" product** when being cited
- As a consequence of this, **versioning mechanisms**, conventions, and uptake differ for software and data
- Where research software is developed to be used by others, it is reasonably common for it to be developed in an **open and transparent** way. In the case of data, it is far more common for it to be **made available (if at all) after it is completed**
- Conversely, there is also software developed and used for **research projects** that is not made available or open during or after development and deployment.
- Major software development **platforms** (e.g., GitHub, GitLab, BitBucket) are widely used in research software development, but these systems are far more commonly used for non-research software development and so function to reflect the needs of a broader group. The tools and platforms used in the data development, publishing and archiving ecosystem are more diverse, and more commonly aligned with research domain needs or interests
- These software development **platforms** serve as development and collaboration platforms for version control systems, but they do not function as archival or **preservation** systems (e.g., the previously but no longer available Gitorious and CodePlex), even if some of the source code is made public this way. Dataset development usually occurs separate from the platforms used for publishing and archiving of that data, and these separate publishing and archiving platforms are better suited for their intended use
- As a consequence of this, software development platforms do not routinely capture **metadata** in the way that a great many data publishing and archiving systems do, using accepted and widely supported minimal metadata standards that used by most research data repositories (e.g., those listed in [re3data.org](re3data.org))
- On the other hand, software packaging systems do capture **metadata**, but are only suitable for research software of a certain type and scale, and which metadata is captured by these systems is determined outside of research needs and interests
- For software, the practice of registering **software in registries** and repositories (e.g., ASCL, RRID, swMath, zenodo) has arisen separately from the platforms used to make that software available, sometimes by users of the software rather than the authors. For data there is no equivalent culture of registering data (especially other's data) separate to the depositing and closely coupled identification of data in a repository system.
- In open source software development, **contributions that are anonymous or from a pseudo-anonymous alias** are more likely to be accepted than in most kinds of research data development

- Data development **roles and authorship recognition** are mature and supported as first class metadata, but software authorship and development roles have not reached this point, especially within version control systems where commits may not match the actual contributions to the software (i.e., some software development roles are not visible on the history of commits, like design, testing and product management)
- **Granularity levels**, as described in (Research Data Alliance/FORCE11 Software Source Code Identification WG et al., 2020), are an aspect of software objects, where a software project can be constituted of multiple software modules and sub-modules that is different from granularity levels for data

These contrasts in current practices represent a different set baseline assumptions for FAIR data and FAIR research software. FAIR principles for research software must address the inherent differences between software and data. They must also recognize that any change in practices sought by a set of FAIR principles for research software must address both a different starting point from data that is not yet FAIR and (potentially) a different end point that for FAIR data.

Previous types of objects, initially papers and then data, use a two-step process that involves an author and a publisher/repository. The author completes the work (the paper or dataset) and then submits it to publisher/repository along with metadata that describes it. The publisher/repository then makes it available (perhaps after some review process). Indexing services can use the metadata and the link to the product to make it findable.

However, software has a long history of being developed collaboratively in version control systems (VCS). It is currently mostly developed using distributed VCS tools (e.g., git and mercurial). When open source software is developed online through VCS platforms (e.g., GitHub, Gitlab, Bitbucket), there is no need for a publisher to accept the software and create an identifier. Similarly, there is no requirement for the developer to record metadata. This leads to there being no permanent identifier (other than a commit hash, debatably), to the metadata being unrecorded, and to the software not being permanently archived.

Note that Software Heritage (Abramatic et al. 2018) is trying to partially fill this gap by archiving software from open source development environments (e.g., GitHub, GitLab, Bitbucket), and capturing intrinsic identifiers for each artifact and metadata that might be available within the source code (in a metadata file like a codemeta.json or CITATION.cff), though this does not address the metadata gap when metadata isn't found in the source code itself. This (Gap 3) and other gaps are further discussed in [Section 5](#).

While another subgroup in the FAIR4RS effort is defining "research software," they are focusing on defining the "research" aspect. We are also concerned with defining "software," and specifically types of software. In Smith et al. (2016), the citation of software is discussed in the content of source code, executables, containers, or virtual machine images, while other software may be available as a service. Katz et al. (2019) starts with this and adds a package as a type of software, and also adds different types of layers as further aspects of software, specifically: infrastructure, libraries, tools, frameworks, and components.

# 3. Translation of the FAIR principles

Having a basic understanding of data and software, we can now begin to evaluate the principles. We start by simply unpacking and translating the FAIR principles (listed in the Appendix) to see how they would apply to research software as written.

The FAIR guiding principles make several references to "(meta)data". These are intended to be expanded as statements about both metadata (which can describe many digital objects) and statements about data, and in doing so highlight the parallels between the suitable handling of data and metadata. We will first consider whether the FAIR guiding principles relating to metadata are, without alteration, suitable for software metadata. Then we consider the FAIR foundational and guiding principles that refer to data and see how suitable they are when swapped for software.

## 3.1 Translation of metadata principles

The guiding principles that refer to "(Meta)data" are given below, altered to refer to metadata alone. These are all suffixed with "a" to indicate that they are narrower variants of the FAIR guiding principles:

> F1a. Metadata are assigned a globally unique and persistent identifier
> F4a. Metadata are registered or indexed in a searchable resource
> A1a. Metadata are retrievable by their identifier using a standardised communications protocol
> A1.1a. The protocol is open, free, and universally implementable
> A1.2a. The protocol allows for an authentication and authorisation procedure, when necessary
> I1a. Metadata use a formal, accessible, shared, and broadly applicable language for knowledge representation.
> I2a. Metadata use vocabularies that follow FAIR principles
> I3a. Metadata include qualified references to other metadata and software*
> R1a. Metadata are richly described with a plurality of accurate and relevant attributes
> R1.1a. Metadata are released with a clear and accessible data usage license
> R1.2a. Metadata are associated with detailed provenance
> R1.3a. Metadata meet domain-relevant community standards

*Note that guiding principle I3a (I3: "*(Meta)data include qualified references to other (meta)data*") is a special case, that the parentheses expand into four relationships. For the purposes of this section we interpret this as "metadata include qualified references to metadata and software." We also consider this a suitable statement about metadata for software, but note that the original guiding principle doesn't reflect this interpretation directly. In the next section we will consider the other expansion: that "software includes qualified references to other metadata and software."

In discussion, no points of disagreement emerged about these guiding principles. The nature of metadata for software (at least as covered by these guiding principles) does not require any alterations. We suggest that these guiding principles regarding metadata should be the same for software metadata as for data metadata. We propose expressing the complete set of guiding principles regarding metadata in F2:

    F2. Software is described with rich metadata (defined by R1 below)

By modifying F2 and adding:

    (defined **first** by R1 below, **and then by the original FAIR principles for metadata**)

## 3.2 Translation of data principles

Continuing our translation, we next simply exchange "(meta)data" for "software" for both the foundational and guiding principles and make minor changes for overall comprehension. As noted in [Section 2](#), "software" here refers to a potentially complex composite of object types. We also apply the principles to objects packaged with the software (e.g., documentation, data) when possible. This gives us the following (all suffixed with "b" to indicate that they are a variant of the original principles):

> F. Findable
> The first step in (re)using software is to find it. Metadata and software should be easy to find for both humans and computers. Machine-readable metadata are essential for automatic discovery of software and services, so this is an essential component of the FAIRification process.
>
> F1b. Software is assigned a globally unique and persistent identifier
> F2b. Software is described with rich metadata (defined by R1 below)
> F3b. Metadata clearly and explicitly include the identifier of the software they describe
> F4b. Software is registered or indexed in a searchable resource
>
> A. Accessible
> Once the user finds the required software, they need to know how it can be accessed, possibly including authentication and authorization.
>
> A1b. Software is retrievable by its identifier using a standardised communications protocol
> A1.1b. The protocol is open, free, and universally implementable

> A1.2b. The protocol allows for an authentication and authorisation procedure, where necessary
> A2b. Metadata are accessible, even when the software is no longer available
>
> I. Interoperable
> The software usually needs to be integrated with other software. In addition, the software needs to interoperate with applications or workflows for analysis, storage, and processing.
>
> I1b. Software uses a formal, accessible, shared, and broadly applicable language for knowledge representation.
> I2b. Software uses vocabularies that follow FAIR principles
> I3b. Software includes qualified references to other metadata and software
>
> R. Reusable
> The ultimate goal of FAIR is to optimise the reuse of software. To achieve this, metadata and software should be well-described so that they can be replicated and/or combined in different settings.
>
> R1b. Software is richly described with a plurality of accurate and relevant attributes
> R1.1b. Software is released with a clear and accessible software usage license
> R1.2b. Software is associated with detailed provenance
> R1.3b. Software meets domain-relevant community standards

As with the previous section, (I3b) is a special case. Here we interpret this principle as making "...reference to metadata and software," and not just "to software."

As mentioned previously, we do not generally discuss the metadata aspects of the principles except when there is a difference from the FAIR data principles. The translated guiding principles F2b, A2b and R1b are also primarily concerned with the metadata about software, but because the original principles explicitly refer to metadata about data we must provide them as translated principles for a new set of principles.

## 3.3 Discussion

Now that we have translated the principles to text that could apply to software, here we:

(1) Discuss the details of the foundational and guiding principles as actually applied to software
(2) Suggest alternate wording where the principle appears inappropriate for software
(3) Try to consider these principles aspirationally, but also point out places where this aspiration has challenges
(4) Propose additional guiding principles modeled on existing ones

*F. Findable*
*The first step in (re)using software is to find it. Metadata and software should be easy to find for both humans and computers. Machine-readable metadata are essential for automatic discovery of software and services, so this is an essential component of the FAIRification process.*

>We believe that findable is an important foundational principle for software.

>We also suggest removing the reference to "services." While software is definitely a component of any service (and a component that should be FAIR), services are considered here an instantiation of software, not the software itself. Services present an additional series of challenges which we have not considered here.

*F1b. Software is assigned a globally unique and persistent identifier*

>This guiding principle is fundamental for any research output, but note that it can take some extra effort from the software creators today to acquire a global and persistent identifier. In [Section 2](), we noted several differences for software development and publishing, both in terms of current practices and in the functionality and existence of relevant infrastructure that might achieve this aim. The creators can use an archive or an institutional repository to keep software and acquire a persistent identifier for their software. However, the identification target might be difficult to choose. As presented in Figure 1 below (from Research Data Alliance/FORCE11 Software Source Code Identification WG et al., 2020), an identification target can be at one of many different granularity levels that are found in a complete software project. For reproducibility for example, it is important to identify a specific version, which means that identifying the full project isn't specific enough. Furthermore there is still a lack of community agreement when it comes to identifying software; see Gaps 1, 2 and 4 in [Section 5]().

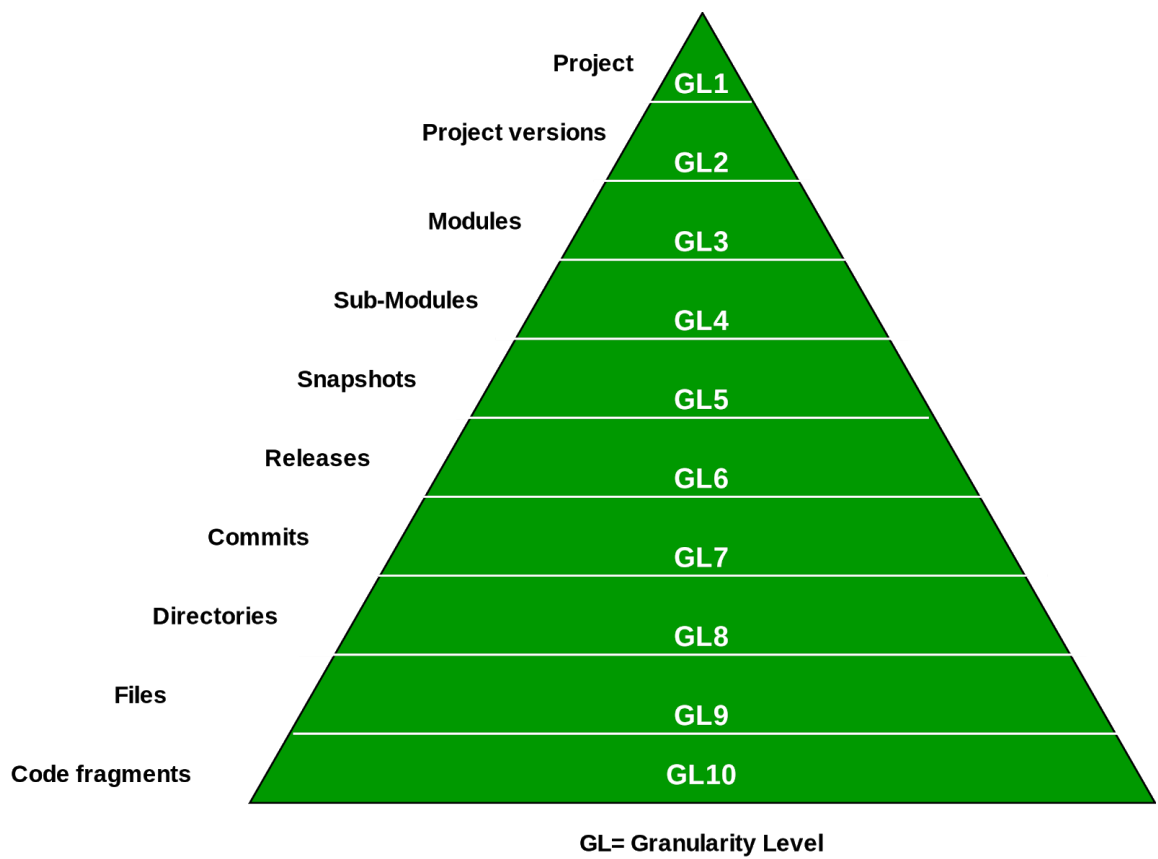

Research Data Alliance/FORCE11 Software Source Code Identification WG et al. (2020). Use cases and identifier schemes for persistent software source code identification (V1.1). *Research Data Alliance*. https://doi.org/10.15497/RDA00053

*F2b. Software is described with rich metadata (defined by R1b below)*

> This guiding principle is reasonable and important when it comes to understanding what the software can do and where it comes from. However, the extent and completeness of the metadata is not yet agreed upon by the research community; see Gaps 1 and 3 in Section 5. As noted in Section 2, software structure can be complex, which adds complexity with the metadata (see Gap 5) and with documentation, which might be considered a metadata element (see Gap 6).
>
> As discussed above, there are several relevant guiding principles that apply without alteration to metadata for digital objects, including software. In order to capture this, we propose changing the wording for this principle to:
>
> "Software is described with rich metadata (defined first by R1b below, and then by the original FAIR principles for metadata)"
>
> The specific principles are F1, F4, A1, A1.1, A1.2, I1, I2, I3, R1, R1.1, R1.2, and R1.3.

*F3b. Metadata clearly and explicitly include the identifier of the software they describe*

> This guiding principle is reasonable. However, there can be many identifiers to different artifacts that are under the same software project; see Gaps 4 and 5 in [Section 5](Section 5).

*F4b. Software is registered or indexed in a searchable resource*

> This guiding principle is reasonable. However, registering software is a complex subject. Current common practice in registries is to identify the software project (see [swMath](swMath), [ASCL](ASCL) or Wikidata) rather than specific software outputs, and this will present a challenge for adopting FAIR software principles; see Gaps 1, 2 and 4 in [Section 5](Section 5). Also see the software structure complexity gap (Gap 5), related to identifiers for different parts of the software.

*A. Accessible*
*Once the user finds the required software, they need to know how it can be accessed, possibly including authentication and authorization.*

> We believe that accessible is an important foundational principle for software.

*A1b. Software is retrievable by its identifier using a standardised communications protocol*

> This guiding principle is reasonable in the abstract, but unclear how to implement it for different types of software, particularly for commercial software. In general, open source software is retrievable by its identifier using a package manager, version control, or similar programmatic download service.

*A1.1b. The protocol is open, free, and universally implementable*

> This guiding principle is reasonable in the abstract, but it is unclear how to implement it for different types of software, particularly for commercial software.

*A1.2b. The protocol allows for an authentication and authorisation procedure, where necessary*

> This guiding principle is reasonable.

*A2b. Metadata are accessible, even when the software is no longer available*

> This guiding principle is reasonable, and some mechanisms for achieving this already exist and are in use for some research software already. For instance, software metadata can be captured in domain specific registries like swMath.org or the Astrophysics Source Code Library (ASCL), in general repository solutions like Zenodo, or via a persistent identifier scheme like DOIs.

Interoperability and Reusability for software

Interoperation and Reuse of data in the original FAIR data principles cover distinct and meaningful categories. Through discussion, we found that the interpretation of these two terms is ambiguous when considering research software. Specifically, both terms are used in the software development context differently than how they are discussed in the FAIR guiding principles (Wilkinson et al., 2016).

To resolve this, we have opted for a narrower interpretation of interoperability, focusing on the mechanisms expressed in software to read, write or otherwise exchange the data or metadata shared between independent software objects. This is compatible with the original intent for interoperability, defined as "the ability of data or **tools** from non-cooperating resources to integrate or **work together** with minimal effort" (boldface added). The original principles enabled this vision primarily through consideration of the form of data and metadata. Here we seek to complement that vision by ensuring the capability exists in software or tools, noting that the paper called not for an exhaustive support for any format of data or metadata, but rather for support for the most general and common mechanisms.

*I. Interoperable*
*Software usually needs to be integrated with other software. In addition, software needs to interoperate with applications or workflows for analysis, storage, and processing.*

> Interoperation between data expresses a reciprocal or concomitant relation. Two data sources can be said to interoperate if they can, with relative ease, be integrated in a way that forms a uniform third object. They are equal contributors to the end result. The potential for integration is commonly taken to be good practice in software engineering, but the nature of that relationship is different. There is a contrast between direct or asymmetrical, and indirect or symmetrical integration.
>
> First, there is the direct and asymmetrical integration between a piece of software and its dependencies. As implied by the label, software becomes dependent on the availability and robustness of those dependencies. The dependencies are integrated into the primary software object. This sense of integration does not seem to reflect the reciprocal relation expressed by "interoperability."
>
> Second, there is an indirect and often symmetrical integration between independent software objects that can or do exchange data. This could be in the form of information passed between two running instances of software (e.g., services), or it could be in the form of support for common data formats read or written by both software packages. This sense of integration does reflect the reciprocal relations expressed by interoperability.

We propose that this foundational principle focus on a sense of interoperability facilitated by the exchange of metadata or data between software following community standards. To better convey this meaning, we propose updating the wording of this foundational principle:

"*Software usually needs to communicate with other software via exchanged data (or possibly its metadata). Software tools can interoperate via common support for the data they exchange.*"

Furthermore, we propose that the sense of direct integration is actually related to the use and reusability of software, rather than interoperability. See the discussion on Reusability foundational principle for more on this point.

As part of this refocus, we will drop some guiding principles that don't reflect this, reword others, and introduce a new principle modelled on one of the reusability guiding principles.

*I1b. Software uses a formal, accessible, shared, and broadly applicable language for knowledge representation.*

This guiding principle applies by default to software source code, which uses a number of different formal, accessible, shared, and broadly applicable languages to represent the knowledge in the software. It may not apply to other representations of the software (e.g. executables, container images, services). Note in particular that for data, this principle serves to discourage a proliferation of ad hoc, informal, undocumented knowledge representations, but because software source code is expressed in a programming language, this is not a problem. Thus, we propose that this is not needed as a guiding principle for software.

*I2b. Software uses vocabularies that follow FAIR principles*

A possible interpretation of this guiding principle is that a software vocabulary is the programming language used to write the software. However, vocabularies in the FAIR data principles refer to ontologies and vocabularies used to model knowledge representation in a given domain. It might be too much of a leap to attach this interpretation to the "vocabularies" term in I2b. Note that most programming languages take many of the steps necessary to be FAIR, but definitely are not reactive to or "follow" the FAIR principles. These programming language "vocabularies" are software projects by themselves and are only visible in the source code form of a software, which means that this guiding principle with this specific interpretation isn't relevant to other software artifacts (e.g., binaries, containers etc.) We decided not to use this interpretation and therefore I2b is not part of our proposed new principles in Section 4.

*I3b. Software includes qualified references to other metadata and software*

This guiding principle applies to software as written, but in discussion we agreed that this is in aid of (re)use of software, rather than interoperability (at least as described above). Additionally, this simple translation of the original guiding principle doesn't capture that qualified references should be to metadata, data and software, as well as to non-digital objects that have a virtual presence in digital systems (e.g., samples, reagents, etc.).

Software source code (and some other types of software) do include references to other software (requirements, imports, libraries, etc.) but not currently in a way that meets F1b and A1b. Software does not generally include references to metadata, though in some cases, it can include (in comments) references to algorithms or other published text that it implements. Some software includes references to external data objects required to execute the software. To be fully FAIR, the data would ideally be FAIR as well, and references to external data fully qualified.

We believe that calling for qualified references to metadata and to data is reasonable. However, in light of the modified definition of the foundational Interoperability principle, we believe that, while the inclusion of guiding principle calling for software to include qualified references to other software is reasonable, this is primarily in aid of the use and reuse of the software. For this reason, we propose that there be two guiding principles:

*"Software includes qualified references to other objects"*
*"Software includes qualified references to other software"*

The second of these is a new guiding principle to be placed under the Reusable foundational principle.

*R. Reusable*
*The ultimate goal of FAIR is to optimize the reuse of software. To achieve this, metadata and software should be well-described so that they can be replicated and/or combined in different settings.*

We believe that useability and reusability is an important foundational principle for software. However, "optimize" is too strong of a statement and should be replaced by "enable and encourage." Finally, software can be described via metadata.

To maximise software (re)use, we must recognise that most software is dependent on other software. FAIR Research Software should be structured to maximise its potential use or reuse. This includes:

- the encapsulation of the software such that it can be reused alone or within other software projects
- the level of abstraction at which the software is expressed

- the record of references to dependencies that enable use and reuse of the software, and
- the metadata that pertains to reusability.

As discussed under the interoperability foundational principle above, it has been difficult to interpret what interoperable means in a FAIR context. This is true for reusable as well. These terms have multiple, overlapping senses when applied to software.

Reuse for software can mean much more than "replicated and/or combined" in the original wording for this foundational principle.

We do not consider executability to be a necessary feature of software for it to be FAIR. There are many legitimate (re)uses of software that do not require executability, for instance, to verify that steps taken within the code are valid, or to look for "bugs" and other errors in the code.

Software is usually written in a human readable form (source code), which will either be executed by an interpreter, or compiled into one or more binary forms suitable for execution within specific hardware and operating system combinations (limiting potential (re)use). We consider making the original human readable form available most harmonious with the FAIR principles, but recognise that for commercial, historical, or sensitivity reasons, the binary or binaries may be the only available form of some software. The binary itself is opaque and may contain bugs and errors. It is impossible to verify its validity and it cannot be modified, for example, to fix bugs. Binaries can be considered black boxes that we can "use" or "reuse" in a research workflow to produce, analyze, or act on data. Source code, on the other hand, can be interrogated, modified, and "reused" in other software or research workflows in a wider range of environments; see Gap 7 in Section 5.

We suggest "replicated, combined, reinterpreted, reimplemented, and/or used" instead of "replicated and/or combined."

A new version of the text above is "The ultimate goal of FAIR is to enable and encourage the use and reuse of software. To achieve this, software should be well-described (by metadata) and appropriately structured so that it can be replicated, combined, reinterpreted, reimplemented, and/or used in different settings."

*R1b. Software is richly described with a plurality of accurate and relevant attributes*

This guiding principle is reasonable.

*R1.1b. Software is released with a clear and accessible software usage license*

This guiding principle is reasonable, assuming that "release" is defined as making the software available. Thus, we think this principle should be re-written as "*Software is made available with a clear and accessible software usage license.*"

*R1.2b. Software is associated with detailed provenance*

This guiding principle is reasonable. A version control system (VCS) may provide detailed provenance for software, but the quality of detail, especially of agents, entities and actions will depend on careful, consistent and considered use of the VCS. Also note that many contributors may not be recorded by a version control system, which by default only stores that single individual who submits each change.

*R1.3b. Software meets domain-relevant community standards*

This guiding principle is reasonable, but requires careful consideration for software, for the reasons in the discussion under the foundational principle and those laid out below.

As noted in [Section 2](), one feature that differentiates software from data is that it is a complex object composed of multiple distinct objects, such as source code and/or binaries, documentation, and possibly data and metadata of various kinds (see Gaps 5 and 6 in [Section 5]() for more discussion). For software, the composition of the complex object may itself be subject to community standards (e.g., an expectation that certain components such as documentation or detailed references to dependencies should be included in the overall object), and the distinct objects may also be subject to separate community standards (i.e., that included or referenced objects should be in a particular form, or otherwise made FAIR in different ways). Software becomes more usable or reusable by meeting these kinds of domain-relevant community standards.

Particularly when considering the source code component of software, community standards may include preferred programming languages or packaging systems. That is, the "domain-relevant community standards" include the norms established around the software community for each programming language. They also include any further norms within research domains. Community standards may include ways of managing and structuring the code, and expectations around the presence and structure of documentation; see Gap 6 in [Section 5](). We interpret this point as allowing multiple domains to operate at once. We do not consider it an aim of the FAIR principles for research software to pursue the integrability of all software with all software or the use of a single preferred programming language above all others.

We also believe that, by extension, this principle can refer to the functionality or capabilities of the software, and that it is reasonable to expect that:

> "Software should read, write or exchange data in a way that meets domain-relevant community standards."

We note that calling for data that is read, written, or exchanged by software to be FAIR would be too strong a statement for data or metadata only used within or between a collection of software. We also do not insist that FAIR software must integrate with repository systems by default (for instance, to capture metadata and issue an identifier); we believe such decisions should be made by the software creator based on how the software will be used, in the context of community standards and expectations.

This interpretation of this principle is harmonious with our proposed interpretation of Interoperability for research software. We propose that this new wording should be a new and separate principle under Interoperability (I1) in addition to preserving the original one as discussed further in [Section 4](#).

## 4. Potential New FAIR Principles for Research Software

In this section, we detail a single set of potential new FAIR principles for research software. In [Appendix B](#), we also compare the potential principles below with a crosswalk table between the GOFAIR FAIR guiding principles and the FAIR4RS subgroup1 FAIR software principles. We use bold text In the crosswalk table to emphasize the differences.

*F. Findable*
*The first step in (re)using software is to find it. Metadata and software should be easy to find for both humans and computers. Machine-readable metadata are essential for automatic discovery of software, so this is an essential component of the FAIRification process.*

*F1. Software is assigned a globally unique and persistent identifier*

*F2. Software is described with rich metadata (defined first by R1 below, and then by the original FAIR principles for metadata)*

*F3. Metadata clearly and explicitly include the identifier of the software they describe*

*F4. Software is registered or indexed in a searchable resource*

*A. Accessible*
*Once the user finds the required software, they need to know how it can be accessed, possibly including authentication and authorization.*

*A1. Software is retrievable by its identifier using a standardised communications protocol*

*A1.1 The protocol is open, free, and universally implementable*

*A1.2 The protocol allows for an authentication and authorisation procedure, where necessary*

*A2. Metadata are accessible, even when the software is no longer available*

*I. Interoperable*

*Software usually needs to communicate with other software via exchanged data (or possibly its metadata). Software tools can interoperate via common support for the data they exchange.*

*I1. Software should read, write or exchange data in a way that meets domain-relevant community standards*

*I2. Software includes qualified references to other objects.*

*R. Reusable*

*The ultimate goal of FAIR is to enable and encourage the use and reuse of software. To achieve this, software should be well-described (by metadata) and appropriately structured so that it can be replicated, combined, reinterpreted, reimplemented, and/or used in different settings..*

*R1. Software is richly described with a plurality of accurate and relevant attributes*

*R1.1. Software is made available with a clear and accessible software usage license*

*R1.2. Software is associated with detailed provenance*

*R1.3. Software meets domain-relevant community standards*

*R2 Software includes qualified references to other software*

## 5. Gaps

In the discussion subsection in [Section 3](), it is clear that there are current gaps that must be overcome to make research software FAIR. To clarify the current situation, we present a non-exhaustive list of gaps that are numbered and named. For each gap, we captured existing or imagined solutions that would resolve the challenges we face when applying the potential FAIR software principles as defined by the FAIR4RS subgroup1 in this report.

Gap 1: metadata and identifier authority

| Gap name | Gap description | Possible solutions |
|---|---|---|
| Metadata and identifier authority | All research software must have unique identifiers and associated metadata. How are these identifiers created? How is the metadata created? **Where are they stored? Who maintains them?** Can we extend metadata schemas so that they describe the inputs and outputs of the software? (Note that metadata may either be extrinsic or intrinsic to the software. Extrinsic metadata can be used to make the software findable, while for intrinsic metadata to be used to make the software findable, something needs to expose it externally. Intrinsic metadata is also guaranteed to be controlled by the authors while extrinsic metadata is controlled by an external authority.) | Note that there are different ways to associate metadata to software in the scholarly ecosystem (while there are other ways to do so in industry, which are not discussed here): |
| | | Capturing the metadata against an identifier, by: |
| | | ● Creating a record in a registry (extrinsic metadata) |
| | | ● Depositing software in an Institutional Repository (IR) or a Zenodo-like repository (extrinsic metadata) |
| | | ● Publishing code with an article with an appropriate publisher (JOSS, IPOL, eLife, etc.) (extrinsic metadata) |
| | | Or keeping the metadata in the original artifact, by: <br> ● Adding metadata in the code itself (intrinsic metadata) |

Gap 2: community agreement on identifiers

| Gap name | Gap description | Possible solutions |
|---|---|---|
| Community agreement on identifiers | At the moment there is no community agreement on what's the best way to identify software for the software authors to get credit.<br>This gap is related to Gap 4 when it comes to choosing the identification target in a software project or identifying the complete software project. | In the case of having a metadata record (in a registry or in a repository), it is "more" FAIR to include the identifier to the software content, which might be a downloadable link to the resource but isn't persistent or a [SWHID](#) to the source code content on the universal software source code archive: Software Heritage. |

Gap 3: community agreement on metadata

| Gap name | Gap description | Possible solutions |
|---|---|---|
| Community agreement on metadata vocabulary and metadata properties | For software, there are vocabularies used by package managers to describe software, but these do not capture metadata about research. There are also discipline-specific vocabularies that sometimes do not capture metadata about software development and usage.<br>At the moment, the research software community hasn't come to an agreement on which vocabulary should be used and which terms improve the FAIRness of the software to satisfy F2. | The [CodeMeta initiative](#) is a subset of the Schema.org vocabulary that was developed to capture properties that aren't available in schema.org. It is also a research community driven project to capture all software metadata vocabularies, available in the CodeMeta [crosswalk](#) table.<br>A possible solution to the gap is to maintain a codemeta.json file inside the software source code repository, which then can be retrieved and indexed in a source code archive.<br>The CodeMeta generator[1] is a tool that can help |

---

[1] https://codemeta.github.io/codemeta-generator/

| | | researchers create a codemeta.json file. |
|---|---|---|

## Gap 4: identification target

| Gap name | Gap description | Possible solutions |
|---|---|---|
| Identification target | What does a software identifier refer to? (for open source software, for commercial software, for a container, for a service, etc.)<br>In addition, there are currently many types of identifiers, and these types are independent and not clearly interoperable.<br>1. Note that every change of a bit in software creates new software, that potentially could have a new identifier and new metadata. (This introduces the idea of a software concept, which is the set of all specific versions of that software.)<br>2. Identifiers, especially those under version control systems, can also remain constant while the source code changes. Examples include repository tags, branches, and access URLs.<br>3. As with the original principles, we do not seek here to mandate a particular frequency for creating identifiers for software, or granularity.<br>4. Are landing pages (Staff et al. 2015) needed for software? If | - Refer to the SCID output for further discussion<br>- These are implementation concerns. We should allow for a diversity in implementations at this stage, rather than a restriction |

| | so, who creates and maintains them? 5. For discussion of these points, see the recent work by the RDA Software Source Code Identification (SCID) working group (Allen et al. 2020) | |

Gap 5: Software structure complexity

| Gap name | Gap description | Possible solutions |
| --- | --- | --- |
| Software complexity | Software is often a complex object made up of other software, data and metadata. How do we deal with this? | - Propose separate sets of principles for each (sub)object not adequately covered by the existing principles, and have a principle within the FAIR software principles that says each part should be handled by the relevant principles. |

Gap 6: Documentation

| Gap name | Gap description | Possible solutions |
| --- | --- | --- |
| Documentation | How do we treat documentation, particularly documentation maintained alongside (or even embedded within) the software, often in the software repository? Do the principles cover it when they refer to metadata? Do they cover it when they refer to software? Do we need to add it as a third type that is | - Documentation standards should be considered and endorsed by domain-relevant community standards groups<br>- Documentation can come in many forms, as a file in the software repository or |

| | explicitly covered by the principles? Is it outside (out of scope) the principles? | as another object, which should be FAIR and archived.<br>- If the documentation is presented on a website, it can be archived on [archive.org](archive.org) and referenced in the metadata of the software.<br>- Another solution for documentation that isn't in the source code is to deposit the documentation separately in a repository (e.g., Zenodo) and reference the documentation PID in the metadata. |
|---|---|---|

Gap 7: Binaries vs. Source Code

| Gap name | Gap description | Possible solutions |
|---|---|---|
| Binaries or services vs. source code | How do we write the principles in a general way, for example so that they apply when only binaries are available and not source code? (This bears in mind that the FAIR principles are not only about open science, open access or open source, but also apply to non-open software.) | Following the discussion in [Section 3.3](Section 3.3), source code is the preferred format for FAIR software, providing a view into the mind of the designer, which is not possible with the binary. With the code, the logic of the software can be "reused" even if the software won't be executed and "used." |

Note that a number of these challenges have also been discussed by Katz et al. (2019) in the context of citation.

# 6. FAIR scope - A perspective of FAIR and beyond FAIR

The FAIR guiding principles for scientific data management and stewardship were intended to drive a change in practice to make a range of Digital Objects increasingly Findable, Accessible, Interoperable and Reusable. Following our vision from the introduction, the FAIR principles for software are a step forward on the path to recognizing software outputs in academia and improving the curation workflows to produce better outputs. Here, we examine the consequences of making software FAIRin practice by reviewing different components of a software output. Additionally, we explore other possible elements or components that are needed to achieve our vision and are beyond the FAIR principles (the original set and the proposed set in this document). For example, executability, robustness, and computational reproducibility are goals we want to achieve but doing so requires more than just the FAIR software principles. Finally, we acknowledge that software is a living and complex object for which it is impossible to propose one solution that fits all software to achieve our vision.

Here we show how a software project can be composed from a collection of modules or packages, where each module/package can contain a set of elements (metadata, documentation, license, source code, executable, dependencies, environment and even an emulated machine). Some elements (e.g., metadata) might be necessary for software to be FAIR. Other elements (open source code or a complete image of the software environment) are desired possibilities, but go beyond the FAIR principles. The most obvious example is having the source code artifacts be accessible and open. The FAIR principles do not require open access, open science, and open source, yet having the source code accessible might give access to intrinsic metadata, documentation, and the software license (which usually is in the source code itself as recommended by the REUSE software project[2]). Note that the figures are simplified and that software can be represented in many more ways with a collection of executables for different environments and also different variants of the source code to produce a specific executable for a specific environment.

The individual figures below (also brought together in Figure 2) show how the addition of certain elements makes software increasingly FAIR and then beyond FAIR, as follows:

- Elements in green are **available** for the software.
- Elements in red are **available but not in a form that supports the proposed vision**.
- The dotted line indicates that the elements can be included or separated from a larger element (for example a license can be available in the source code or in a separate location. The same is possible with the environment, it can be packaged as a Docker image or accessed in different locations).
- The museum icon indicates that the software is preserved in a secured location with the purpose of archiving (e.g., Zenodo, Software Heritage, etc.)

---

[2] https://reuse.software/

- The cog icon indicates that the software can be easily executed and might answer the reproducibility challenge.

| | |
|---|---|
| 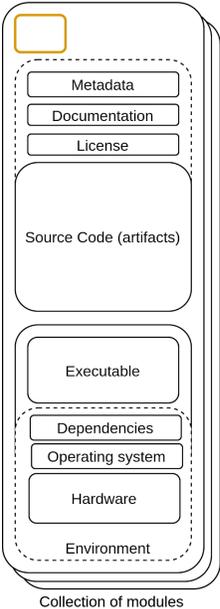 Collection of modules (unFAIR Software diagram) | **Type 1 - unFAIR software:**<br>Software where there are **elements (if available) that are not immediately findable or accessible** This can happen when software is simply named in an article. When this occurs, it can be difficult to find the correct software, let alone access it. |
| 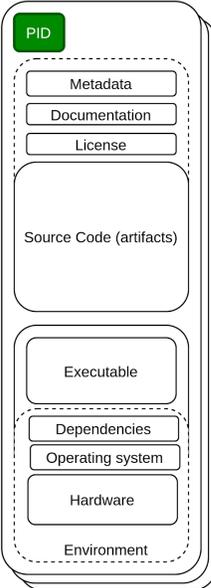 Identifiable (with PID) | **Type 2 - Identifiable:**<br>Software with only a PID element available, nothing else. This may **unambiguously identify** the software (which is better than only a name mentioned), but it doesn't mean that the software or rich metadata are available. For example, this might be a registry entry with the name of the software and no additional metadata. |

| 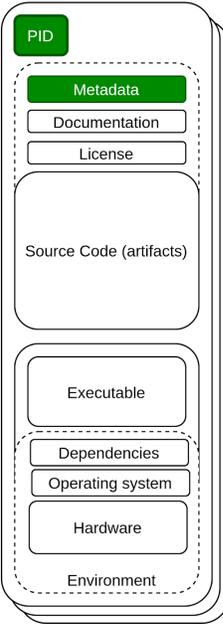 FAIR Metadata | **Type 3 - FAIR metadata:**<br>Software with available **PID and metadata** elements; this is the case of a metadata record in a registry, usually with a PID registered in the specific registry (e.g., ASCL, swMath, RRID etc.) |
|---|---|
| 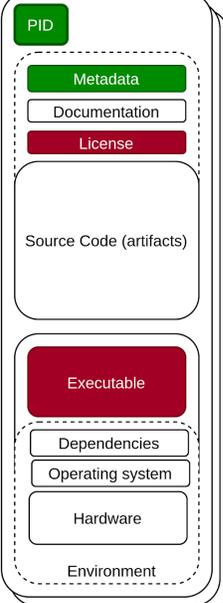 FAIR Metadata / Restricted Access (authentification) | **Type 4 - FAIR metadata/restricted access (authentification):**<br>Software with a **PID, metadata** and an **executable** under **restrictive usage** (with the appropriate restrictive/proprietary license). This is the case of proprietary software that has a metadata record in a registry and permits access to the software executable to some limited set of people, potentially who buy it or are members of a specific permitted group (e.g., software in an organization that is only open to the active members of the organization). This could be considered FAIR software if some other requirements are met *(FAIR vocabulary for the metadata, follow community standards, standardised communications protocol which is open, free and implementable, detailed provenance, etc.)* |
| 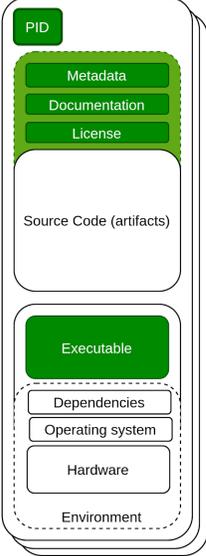<br>FAIR Software / Full access to Software executable | **Type 5 - FAIR software / full access to software executable:**<br>Software with a **PID, metadata, documentation and an executable**. By having the documentation (e.g., a manual or other instructions) the reusability of the software might be slightly better than in the fourth box. This is the first box that shows the minimal elements for a FAIR piece of software.<br>An executable is a file that can be executed by a specific environment (can be a .exe for Microsoft, a .deb for a Linux Debian distribution, etc.). This file (or files) is like a black box, it can't be verified or modified to resolve errors. Having only an executable is not a very robust way to keep software. |
| 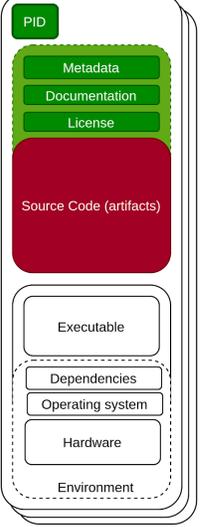<br>FAIR Software / Full access to source code on dev. platform | **Type 6 - FAIR software / full access to the source code on a development platform:**<br>Software with a **PID, metadata, documentation, license and the source code,** which are available on a **software development platform.** Having the source code is a good step forward in providing access to the knowledge in the software. Despite this, source code that is not archived can disappear with the platform or if the author decides to delete the repository. |
| 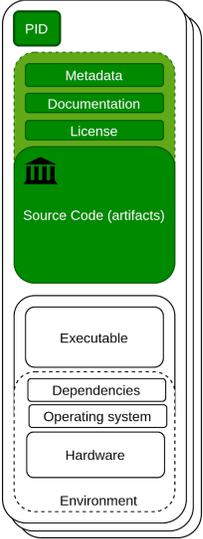 FAIR software and Open Source code archived | **Type 7 - FAIR software and Open Source code archived:** An archived Open Source piece of software including a **PID, metadata, documentation, license and the source code.** Note that it is easier for software to be FAIR when it is open source, has the metadata, license and documentation included in the source code and has an intrinsic identifier (e.g., a SWHID) because open source software usually lives in a public repository that can be already saved in Software Heritage or a save request can be submitted at any time by anyone[3]. |
| 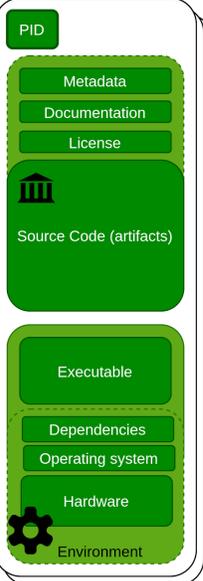 FAIR software, Open Source and Reproducible | **Type 8 - FAIR software, Open source and reproducible:** The holy grail of software reproducibility, showing an exhaustive set of elements included with software: **PID, metadata, documentation, license, source code, executable(s), environment, dependencies and hardware** (which can be emulated or physical hardware). |

---

[3] https://archive.softwareheritage.org/save/

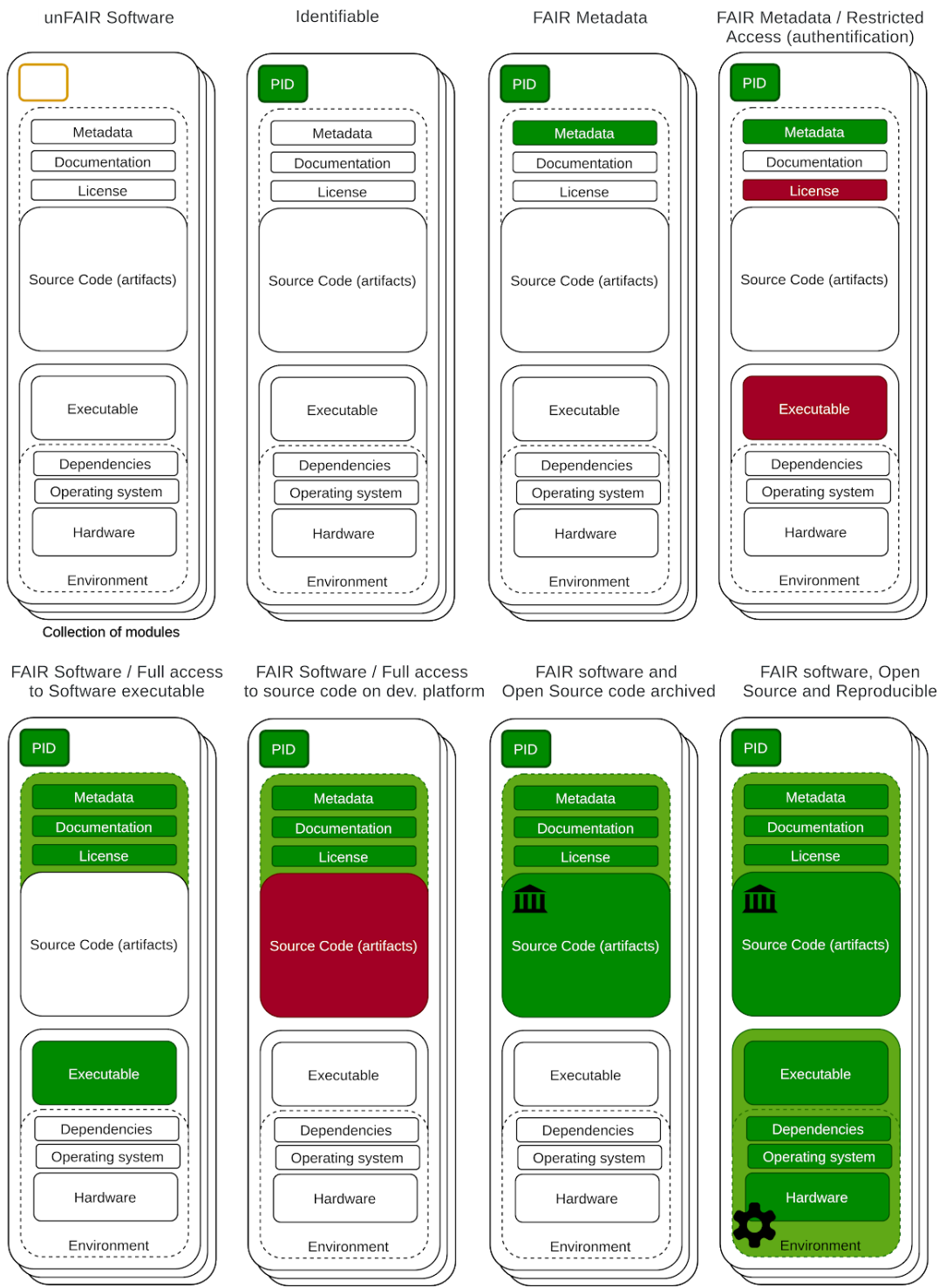

Figure 2: Summarizing software as increasingly FAIR research objects (inspired by the FORCE11 diagram[4])

---
[4] https://www.force11.org/fairprinciples

# Appendix A: the FAIR guiding principles for data

We base our work on the FAIR principles (Wilkinson et al. 2016) as listed by the GO FAIR initiative (https://www.go-fair.org/fair-principles/) as of 19 August 2020, specifically:

F. Findable
The first step in (re)using data is to find them. Metadata and data should be easy to find for both humans and computers. Machine-readable metadata are essential for automatic discovery of datasets and services, so this is an essential component of the FAIRification process.

F1. (Meta)data are assigned a globally unique and persistent identifier
F2. Data are described with rich metadata (defined by R1 below)
F3. Metadata clearly and explicitly include the identifier of the data they describe
F4. (Meta)data are registered or indexed in a searchable resource

A. Accessible
Once the user finds the required data, she/he needs to know how can they be accessed, possibly including authentication and authorisation.

A1. (Meta)data are retrievable by their identifier using a standardised communications protocol
A1.1 The protocol is open, free, and universally implementable
A1.2 The protocol allows for an authentication and authorisation procedure, where necessary
A2. Metadata are accessible, even when the data are no longer available

I. Interoperable
The data usually need to be integrated with other data. In addition, the data need to interoperate with applications or workflows for analysis, storage, and processing.

I1. (Meta)data use a formal, accessible, shared, and broadly applicable language for knowledge representation.
I2. (Meta)data use vocabularies that follow FAIR principles
I3. (Meta)data include qualified references to other (meta)data

R. Reusable
The ultimate goal of FAIR is to optimise the reuse of data. To achieve this, metadata and data should be well-described so that they can be replicated and/or combined in different settings.

R1. (Meta)data are richly described with a plurality of accurate and relevant attributes
R1.1. (Meta)data are released with a clear and accessible data usage license
R1.2. (Meta)data are associated with detailed provenance
R1.3. (Meta)data meet domain-relevant community standards

# Appendix B: The FAIR data principles compared to the proposed FAIR software principles

| FAIR principles (Wilkinson et al. 2016) as listed by GO FAIR | *FAIR software principles as listed in [Section 4](Section 4) (changes are in **bold**)* |
|---|---|
| F. Findable<br>The first step in (re)using data is to find them. Metadata and data should be easy to find for both humans and computers. Machine-readable metadata are essential for automatic discovery of datasets and services, so this is an essential component of the FAIRification process. | *F. Findable*<br>*The first step in (re)using **software** is to find **it**. Metadata and **software** should be easy to find for both humans and computers. Machine-readable metadata are essential for automatic discovery of **software**, so this is an essential component of the FAIRification process.* |
| F1. (Meta)data are assigned a globally unique and persistent identifier | *F1. **Software is** assigned a globally unique and persistent identifier* |
| F2. Data are described with rich metadata (defined by R1 below) | *F2. **Software is** described with rich metadata (defined **first** by R1 below, **and then by the original FAIR principles for metadata**)* |
| F3. Metadata clearly and explicitly include the identifier of the data they describe | *F3. Metadata clearly and explicitly include the identifier of the **software** they describe* |
| F4. (Meta)data are registered or indexed in a searchable resource | *F4. **Software is** registered or indexed in a searchable resource* |
| A. Accessible<br>Once the user finds the required data, she/he needs to know how can they be accessed, possibly including authentication and authorisation. | *A. Accessible*<br>*Once the user finds the required **software**, **they** need to know how **it** can be accessed, possibly including authentication and authorization.* |
| A1. (Meta)data are retrievable by their identifier using a standardised communications protocol | *A1. **Software is** retrievable by **its** identifier using a standardised communications protocol* |
| A1.1 The protocol is open, free, and universally implementable | *A1.1 The protocol is open, free, and universally implementable* |
| A1.2 The protocol allows for an authentication and authorisation procedure, where necessary | *A1.2 The protocol allows for an authentication and authorisation procedure, where necessary* |
| A2. Metadata are accessible, even when the data are no longer available | *A2. Metadata are accessible, even when the **software is** no longer available* |
| I. Interoperable<br>The data usually need to be integrated with other data. In addition, the data need to interoperate with applications or workflows for analysis, storage, and processing. | *I. Interoperable*<br>*S**oftware** usually needs to **communicate** with other **software** via **exchanged data (or possibly its metadata)**. Software tools can interoperate via common support for the data they exchange.* |
| I1. (Meta)data use a formal, accessible, shared, and broadly applicable language for knowledge representation. | **(deemed unnecessary)** |
| I2. (Meta)data use vocabularies that follow FAIR principles | **(deemed unnecessary)** |
| *R1.3. [(Meta)data meet domain-relevant community standards] used as a model for a new Interoperability guiding principle* | **I1. Software should read, write or exchange data in a way that meets domain-relevant community standards** |
| I3. (Meta)data include qualified references to other (meta)data | *I2. **Software** includes qualified references to other **objects**.* |
| R. Reusable<br>The ultimate goal of FAIR is to optimise the reuse of data. To achieve this, metadata and data should be well-described so that they can be replicated and/or combined in different settings. | *R. Reusable*<br>*The ultimate goal of FAIR is to **enable and encourage** the **use and** reuse of **software**. To achieve this, **software** should be well-described **(by metadata) and appropriately structured** so that it can be replicated, combined, **reinterpreted, reimplemented, and/or used** in different settings.* |
| R1. (Meta)data are richly described with a plurality of accurate and relevant attributes | *R1. **Software is** richly described with a plurality of accurate and relevant attributes* |
| R1.1. (Meta)data are released with a clear and accessible data usage license | *R1.1. **Software is** made available with a clear and accessible **software** usage license* |
| R1.2. (Meta)data are associated with detailed provenance | *R1.2. **Software is** associated with detailed provenance* |
| R1.3. (Meta)data meet domain-relevant community standards | *R1.3. **Software** meets domain-relevant community standards* |
| *I3. [(Meta)data include qualified references to other (meta)data] used as a model for a new Reusability guiding principle* | **R2 Software includes qualified references to other software** |

# Appendix C: FAIR4RS Subgroup 1 Membership

The full list of members of Subgroup 1 of the FAIR for Research Software (FAIR4RS) FORCE11 working group, RDA working group, and ReSA task force are:

- Mawada Ali, CBSB Sudan
- Hartwig Anzt, University of Tennessee / Karlsruhe Institute of Technology
- Christian Busse, Free Software Foundation Europe
- Marie-Josée Cros, INRAE, France
- Piotr Wojciech Dabrowski, HTW Berlin University of Applied Science
- Carole Goble, University of Manchester
- Charles Gray, Newcastle University
- Morane Gruenpeter, Software Heritage, Inria
- Maggie Hellström, Lund University, Sweden and ICOS Carbon Portal
- Tom Honeyman, Australian Research Data Commons
- Lorraine Hwang, UC Davis, CIG
- Catherine Jones, STFC
- Daniel S. Katz, University of Illinois at Urbana-Champaign
- Adriaan Klinkenberg, Elsevier (SoftwareX, Computer Physics Communications)
- Matthias Liffers, Australian Research Data Commons
- Nick Lynch, Curlew Research
- Paula Andrea Martinez, The University of Queensland / National Imaging Facility
- Sergio Martinez Cuesta, University of Cambridge and AstraZeneca
- Javier Moldon, IAA-CSIC
- Daniel Nüst, University of Münster | de-RSE
- Manodeep Sinha, Swinburne University of Technology
- Vanessa Sochat, Stanford University
- Sarah Stevens, University of Wisconsin-Madison
- Ilian Todorov, UKRI STFC
- Mark D. Wilkinson, CBGP UPM-INIA, Madrid
- Yo Yehudi, Wellcome Trust